\documentclass[pdflatex,sn-mathphys-num]{sn-jnl}

\usepackage[T1]{fontenc}
\usepackage[utf8]{inputenc}

\usepackage{lmodern,amssymb,amsbsy,amsmath,bm,slashed,simpler-wick,hypernat,hyperref}

\usepackage[caption=false,labelfont=normalsize,labelformat=empty,textfont=normalsize, justification=centering]{subfig}

\renewcommand{\Im}{\mathrm{Im}}
\newcommand{\D}{\mathrm{d}}
\newcommand{\iu}{\mathrm{i}}

\usepackage{layouts} 

\begin{document}

\title[Unitarity, the optical theorem, and the Pauli exclusion principle]{Unitarity, the optical theorem, and the Pauli exclusion principle}

\author*[1]{\fnm{Peter} \sur{Mat\'ak}}\email{peter.matak@fmph.uniba.sk}

\affil*[1]{\orgdiv{Department of Theoretical Physics}, \orgname{Faculty of Mathematics, Physics and Informatics, Comenius University in Bratislava}, \orgaddress{\street{Mlynsk\'a dolina}, \city{Bratislava}, \postcode{84248}, \country{Slovak Republic}}}

\abstract{
We show that the fermionic exclusion principle in scattering problems manifests itself through constraints implied by unitarity and the optical theorem. Configurations that formally allow identical fermions to appear in the same quantum state at the level of intermediate amplitudes are not pathological. Instead, they turn out to be essential for implementing the Pauli principle in scattering processes. Making this connection explicit resolves an apparent tension between the exclusion principle and unitarity and provides a clarified view of how fermionic statistics manifests within the $S$‑matrix framework.
}


\maketitle

\section{Introduction}\label{sec1}

In perturbative quantum field theory, the optical theorem follows directly from the $S$-matrix unitarity~\cite{Heisenberg:1943hhj}. Writing $S = 1 + \iu T$ in the unitarity condition $S^{\dagger} S = 1$ and taking the diagonal matrix elements, we obtain
\begin{align}\label{eq1}
2\Im T^{\vphantom{\dagger}}_{ii}=\sum_f\vert T^{\vphantom{\dagger}}_{fi}\vert^2
\end{align}
relating the imaginary part of the forward scattering amplitude on the left-hand side to the sum of squared amplitudes for all $i \to f$ reactions, integrated over the final-state momenta. Following standard conventions, we further write
\begin{align}\label{eq2}
T_{fi} = (2\pi)^4 \delta^{(4)}(p_f - p_i) M_{fi}
\end{align}
where $p_f$ and $p_i$ denote the four-momenta of the final and initial state particles, respectively.

\begin{figure}
\subfloat{\label{fig1a}}
\subfloat{\label{fig1b}}
\subfloat{\label{fig1c}}
\centering\includegraphics[scale=1]{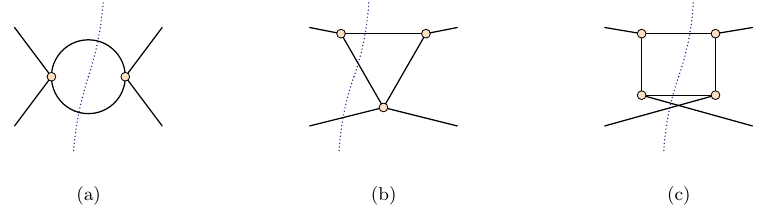}
\caption{\label{fig1} Example topologies of diagrams entering the right-hand side of \eqref{eq1}. The cut diagram in Fig.~\ref{fig1b} represents the anomalous-threshold contribution \cite{Hannesdottir:2022bmo}. In this paper, we study diagrams similar to Fig.~\ref{fig1c} in which the crossed legs correspond to a fermionic particle.}
\end{figure}

With the exception of non-Abelian gauge field theories, a stronger version of the optical theorem can be formulated for individual diagrams~\cite{Cornwall:2010upa}. For a forward-scattering diagram with a nonzero imaginary part represented by an on-shell cut, we can thus identify a corresponding contribution to the transition probability associated with the specific process whose final state is defined by that cut. Typical textbook examples \cite{Peskin:1995ev, Zee:2003mt, Schwartz:2014sze, Cornwall:2010upa, Horejsi:2026kaf} then include cuts of self-energy or scattering loop diagrams into two connected parts, as illustrated in Fig.~\ref{fig1a}. 

However, a connected forward-scattering diagram does not necessarily imply the connectedness of the components to which it is cut, and two peculiar cases may occur. The first of these, an example of which is shown in Fig.~\ref{fig1b}, arises from an interference between a connected and a disconnected diagram. These contributions are related to anomalous thresholds, which, for two-particle initial states, occur only if at least one of the particles involved is unstable~\cite{Hannesdottir:2022bmo}. Within the framework of kinetic theory and the Boltzmann equation, anomalous thresholds have been shown to approximate thermal-mass effects in lower-order reaction rates~\cite{Blazek:2021gmw, Blazek:2022azr}. 

Finally, the crossed-leg diagram in Fig.~\ref{fig1c} is cut into two disconnected parts. In the literature, such contributions are relatively rare and are typically discussed only in specific contexts. In Ref.~\cite{Frye:2018xjj}, for example, a diagram with crossed photonic legs was included to obtain an infrared-finite cross section (see (20) in Ref.~\cite{Frye:2018xjj}). In Ref.~\cite{Blazek:2021zoj}, the physical interpretation of crossed-leg diagrams has been discussed in terms of quantum statistical effects.

In the present work, we focus on the simplest realization of the crossed-leg contribution involving fermionic particles, in which even the forward-scattering diagram, whose imaginary part we compute on the left-hand side of \eqref{eq1}, is a tree-level diagram. Unlike in Ref.~\cite{Blazek:2021zoj}, we do not take the density-matrix evolution as a starting point; instead, using a simple toy model, we analyze a hypothetical collider experiment. In doing so, we believe that the reader will gain a clearer understanding of these effects within the framework of perturbative scattering theory.

The paper is structured as follows. Section~\ref{sec2} illustrates an apparent violation of the Pauli exclusion principle in a cut of a crossed-leg forward-scattering diagram. Section~\ref{sec3} explains how the interference between two disconnected amplitudes resolves the apparent paradox. Finally, section~\ref{sec4} provides a summary.

\section{Two fermions occupying the same quantum state?}\label{sec2}

\begin{figure}
\subfloat{\label{fig2a}}
\subfloat{\label{fig2b}}
\subfloat{\label{fig2c}}
\centering\includegraphics[scale=1]{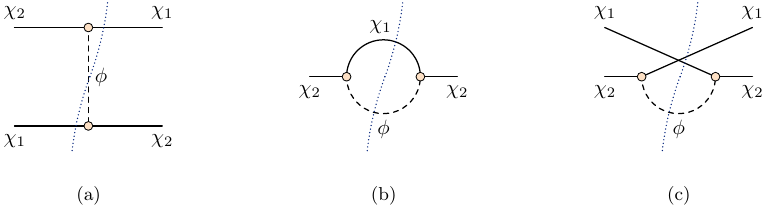}
\caption{\label{fig2} The lowest-order connected forward-scattering diagram for the $\chi_1\chi_2$ initial state within the model defined in \eqref{eq3} (diagram (a)). Diagrams (b) and (c) correspond to \eqref{eq10} and \eqref{eq13} integrated over the full phase-space, respectively.}
\end{figure}

Let us consider a model in which two Majorana fermions, $\chi_1$ and $\chi_2$, with respective masses $m_1$ and $m_2$, interact via
\begin{align}\label{eq3}
L_{\text{int}} = -g\bar{\chi}_2\chi_1\phi
\end{align}
where $\phi$ is a light neutral scalar of mass $m_\phi$. Considering an experiment in which a beam of $\chi_1$ hits a target consisting of $\chi_2$ particles, what final states can be expected at lowest order in the coupling constant $g$? With \eqref{eq1} in mind, this question can be addressed by drawing forward-scattering diagrams and identifying possible on-shell cuts representing the imaginary part. Assuming $m_2 > m_1+m_\phi$ for the masses, the forward scattering diagram in Fig.~\ref{fig2a} can already be made from just two vertices. The diagram comes with a cut and represents a contribution to a transition probability. If treated as a contribution to an amplitude, squared and integrated over the final-state momenta, it would lead to a $t$-channel singularity, as discussed in Refs.~\cite{Peierls:1961zz, Melnikov:1996na, Melnikov:1996iu, Grzadkowski:2021kgi, Karamitros:2022nnh}. The imaginary part of this diagram can be straightforwardly extracted using
\begin{align}\label{eq4}
\Im\,\frac{1}{(p_2-p_1)^2-m^2_\phi+\iu\epsilon}=-\pi\delta[(p_2-p_1)^2-m^2_\phi]
\end{align}
for the scalar propagator. Then we obtain
\begin{align}\label{eq5}
2\Im\, M_{\chi_1\chi_2\to\chi_1\chi_2} = -2\pi\delta[(p_2-p_1)^2-m^2_\phi] \times \vert M(p_1,p_2)\vert^2
\end{align}
with
\begin{align}\label{eq6}
\iu M(p_1,p_2) = \iu g\,\bar{u}(p_1,s_1)u(p_2,s_2)
\end{align}
and $p_1$, $p_2$ and $s_1$, $s_2$ denoting the four-momenta and spin projections of the initial-state particles, respectively.

Nevertheless, we may still hesitate to treat the diagram in Fig.~\ref{fig2a} as a valid contribution. According to \eqref{eq1}, it represents a squared amplitude of the $\chi_1\chi_2 \to \chi_1\chi_1\phi$ process, where the final state consists of particles whose lines are cut. However, the two $\chi_1$ particles in the final state appear in the same quantum state. By the Pauli exclusion principle, such a contribution might be expected to vanish identically, but, as we see in \eqref{eq5}, it does not. The diagram shown in Fig.~\ref{fig2a} thus challenges our understanding of the optical theorem and motivates us to seek its physical interpretation.

\section{Interference of two disconnected amplitudes}\label{sec3}

To better understand the specific situation, it may be helpful to take a closer look at how the diagram in Fig.~\ref{fig2a} contributes to the transition probability in the optical theorem, i.e. the right-hand side of \eqref{eq1}. 

In the tree-level amplitude for the $\chi_1\chi_2 \to \chi_1\chi_1\phi$ reaction, let $\bm{p}_1$ and $\bm{p}_2$ denote the three-momenta of the initial-state particles. In the final state, we have $\bm{k}$ and $\bm{k}'$ for the two $\chi_1$ particles, with spin projections $r$ and $r'$, and a three-momentum $\bm{k}_\phi$ for the scalar. With a single insertion of the Lagrangian density from \eqref{eq3} we find two possible ways to contract the $\chi_1$ field with the external states. These are
\begin{align}\label{eq7}
\langle 0\vert \wick{
\c2 a^{\smash{\vphantom{\dagger}}}_{\bm{k}\smash{'}} 
\c1 a^{\smash{\vphantom{\dagger}}}_{\bm{k}} \ldots \c1 \chi_1\ldots 
\c2 a^{\smash{\dagger}}_{\smash{\bm{p}_1}}
}\vert 0\rangle\quad\text{and}\quad
\langle 0\vert \wick{
\c2 a^{\smash{\vphantom{\dagger}}}_{\bm{k}\smash{'}} 
\c1 a^{\smash{\vphantom{\dagger}}}_{\bm{k}} \ldots \c2 \chi_1\ldots 
\c1 a^{\smash{\dagger}}_{\smash{\bm{p}_1}}
}\vert 0\rangle
\end{align}
and lead to
\begin{align}
\iu M_1=&\; 2E_{\bm{p}_1}(2\pi)^3\delta_{s_1r'}\delta^{(3)}(\bm{p}_1-\bm{k}')\times 
\iu M(k,p_2)\label{eq8}\\
\iu M_2=&\; 2E_{\bm{p}_1}(2\pi)^3\delta_{s_1r\hphantom{'}}\delta^{(3)}(\bm{p}_1-\bm{k})\times
-\iu M(k',p_2)\label{eq9}
\end{align}
differing only by the role of $\bm{k}$, $r$ and $\bm{k}'$, $r'$ and a relative minus sign.

When computing the transition probability from \eqref{eq8} and \eqref{eq9}, we observe that $\vert M_1\vert^2$ and $\vert M_2\vert^2$ contribute equally, and the same holds for the interference terms $M^{\vphantom{*}}_{\smash{1}} M^*_{\smash{2}}$ and $M^{\vphantom{*}}_{\smash{2}} M^*_{\smash{1}}$. We therefore replace $\vert M^{\vphantom{*}}_{\smash{1}} + M^{\vphantom{*}}_{\smash{2}}\vert^2$ by $2M^{\vphantom{*}}_{\smash{1}}(M^*_{\smash{1}}+M^*_{\smash{2}})$ and obtain two types of contributions. The first part comes from $2\vert M_1\vert^2$. Accounting for the proper normalization of all external states, it yields
\begin{align}
&\frac{1}{2E_{\bm{p}_1}V_3}\frac{1}{2E_{\bm{p}_2}V_3}V_4(2\pi)^4\vert M(k,p_2)\vert^2
\D\mathrm{\Phi}_2(p_2;k,k_\phi)\label{eq10}\\
&\times(2E_{\bm{p}_1})^2V_3(2\pi)^3\delta_{s_1r'} \delta^{(3)}(\bm{p}_1-\bm{k}')
\frac{\D^3\bm{k}'}{(2\pi)^32E_{\bm{k}'}}\nonumber
\end{align}
where the factor of two has canceled out the symmetry factor of one half due to the two identical particles in the final state. The two-particle phase-space element is introduced as
\begin{align}\label{eq11}
\D\mathrm{\Phi}_2(p_2;k,k_\phi)=\delta^{(4)}(p_2-k-k_\phi)\frac{\D^3\bm{k}}{(2\pi)^32E_{\bm{k}}}
\frac{\D^3\bm{k}_\phi}{(2\pi)^32E_{\bm{k}_\phi}}
\end{align}
and $V_n = (2\pi)^n \delta^{(n)}(0)$. Integrating \eqref{eq10} over $\D^3\bm{k}'$, summing over $r'$, and dividing by the $V_4/V_3$ ratio, we obtain the transition probability per unit time, or
\begin{align}\label{eq12}
\D\mathrm{\Gamma}=\frac{1}{2E_{\bm{p}_2}}(2\pi)^4\vert M(k,p_2)\vert^2\D\mathrm{\Phi}_2(p_2; k, k_\phi) 
\end{align}
which represents the differential width for the $\chi_2 \to \chi_1 \phi$ decay. This result reflects a trivial situation in which the decaying $\chi_2$ remains unaffected, while the $\chi_1$ particle passes by inconspicuously.

The second contribution, originating from the $2M^{\vphantom{*}}_{\smash{1}} M^*_{\smash{2}}$, is more interesting. It takes the form
\begin{align}
&-\frac{1}{2E_{\bm{p}_1}V_3}\frac{1}{2E_{\bm{p}_2}V_3}(2\pi)^8 M(k,p_2)M(k',p_2)^*
\delta^{(4)}(p_2-k'-k_\phi)\D\mathrm{\Phi}_2(p_2;k,k_\phi)
\vphantom{\frac{\D^3\bm{k}'}{(2\pi)^32E_{\bm{k}'}}}\label{eq13}\\
&\times 2E_{\bm{p}_1}(2\pi)^3\delta_{s_1r}\delta^{(3)}(\bm{p}_1-\bm{k}) 2E_{\bm{p}_1}(2\pi)^3\delta_{s_1r'}\delta^{(3)}(\bm{p}_1-\bm{k}')\frac{\D^3\bm{k}'}{(2\pi)^32E_{\bm{k}'}}\nonumber
\end{align}
and repeating the same steps that led to \eqref{eq12} we find
\begin{align}\label{eq14}
-\frac{(2\pi)^3\delta_{s_1r}\delta^{(3)}(\bm{p}_1-\bm{k})}{V_3}\times\D\mathrm{\Gamma}
\end{align}
having a straightforward interpretation. The contribution in \eqref{eq14} arises only if the momentum and spin of the $\chi_1$ produced in the decay match those of the $\chi_1$ already present in the beam. In that case it cancels the result in \eqref{eq12} in full agreement with the Pauli exclusion principle.

We can now integrate over the entire final-state phase space in \eqref{eq10} and \eqref{eq13} with the results illustrated by the cut diagrams in Figs.~\ref{fig2b} and \ref{fig2c}, respectively. In Fig.~\ref{fig2b}, we may imagine an additional $\chi_1$ line contributing a factor of $2E_{\bm{p}_1}V_3$ canceled by the state normalization in finite volume. The diagrams in Figs.~\ref{fig2a} and \ref{fig2c} are identical, and, up to normalization factors, the phase-space integration in \eqref{eq14} reproduces \eqref{eq5}. 

In principle, it is possible to express the large three‑dimensional volume in the denominator of \eqref{eq14} in terms of the incoming particle flux and define a cross section, treating the Pauli‑blocking effect as a two‑particle process. However, the resulting cross section would be negative and would contain a delta function. It is important to keep in mind that similar contributions cannot be interpreted as simple products of two diagrams, as might seem to be the case from the right-hand side of \eqref{eq1}. If crossed fermionic legs are cut, there is always an additional minus sign, as can be understood from the above discussion of the two interfering contributions. Such a diagram can never appear on its own and only makes sense when combined with the one in Fig.~\ref{fig2b}.

\section{Summary}\label{sec4}

In this paper, we studied the interplay between two seemingly unrelated concepts--the Pauli exclusion principle and the optical theorem. Using a simple toy model with two Majorana fermions interacting via a light scalar mediator, we showed that, for specific initial states, the optical theorem implies contributions in which two fermions appear to occupy the same quantum state--a scenario that seems to violate the Pauli exclusion principle. We then demonstrated that such terms naturally arise from the interference of two disconnected diagrams. A full calculation confirms that the exclusion principle is not violated, with contributions from same-state fermions canceling exactly. Or, in other words, the optical theorem remains valid since we have shown that it is fully in agreement with the Pauli exclusion principle.

\backmatter

\bmhead{Acknowledgements}

The author thanks Tom\'{a}\v{s} Bla\v{z}ek for his careful reading of the manuscript and valuable feedback. This work was supported by the Slovak Grant Agency VEGA, project No. 1/0162/26, and by the Slovak Ministry of Education under Contract No. 0466/2022.

\bibliography{CLANOK.bib}

\end{document}